\documentclass[a4paper,11pt]{article}
\usepackage{jheppub} 
\usepackage{orcidlink}
\usepackage{dsfont}

\newcommand{\Scal}{\mathcal{S}}

\newcommand{\Ncal}{\mathcal{N}}
\newcommand{\Mcal}{\mathcal{M}}

\newcommand{\Dcal}{\mathcal{D}}

\newcommand{\Vcal}{\mathcal{V}}

\newcommand{\mI}{m_I^{D-2}}
\newcommand{\id}{\mathds{1}}

\title{\boldmath Exact AdS and pp-waves in multi-gravity beyond pairwise interactions}

\author{J. Flinckman\orcidlink{0009-0004-4545-3123}}
\affiliation{Department of Physics \& The Oskar Klein Centre,\\
Stockholm University, AlbaNova University Centre, SE-106 91 Stockholm, Sweden}

\emailAdd{Joakim.flinckman@fysik.su.se}

\abstract{We construct exact anti-de Sitter (AdS) waves and pp-waves in ghost-free multi-gravity beyond pairwise interactions for an arbitrary number $\mathcal N$ of vielbeins, in any spacetime dimension $D\geq3$. Deforming proportional AdS and Minkowski solutions along a common null direction reduces the full nonlinear field equations to coupled linear differential equations for the wave profiles, without restricting their amplitudes. The system is diagonalised in the same eigenbasis as the perturbative Fierz–Pauli mass matrix, separating the solutions into one massless and $\mathcal N-1$ massive eigenprofiles. We obtain explicit radial and transverse solutions and impose an AdS boundary falloff. This falloff admits the appropriate radial branch of every mode with non-negative squared mass and excludes profiles with negative squared masses, including logarithmic solutions at the Breitenlohner–Freedman threshold. We also relate the AdS waves to pp-waves through a local flat limit. Explicit examples exhibit wave configurations that cannot be reproduced by ghost-free pairwise theories, demonstrating that genuine multi-field interactions support new exact gravitational-wave solutions.}

\begin{document}
\maketitle
\flushbottom

\section{Introduction}
\label{sec:intro}
General Relativity describes the gravitational interaction through a single dynamical spacetime metric. In the linearised theory, gravitational waves correspond to excitations of a massless spin-2 field. This connection between geometry and field theory raises the question of whether gravity can consistently involve more than one spin-2 field, in analogy with the multiple spin-1 fields of Yang–Mills theory. While spin-1 interactions are central to the Standard Model, interactions between spin-2 fields are much less explored and understood.

A major obstacle to such constructions is the presence of Boulware--Deser ghosts, pathological degrees of freedom that generically appear in nonlinear theories of interacting spin-2 fields \cite{Boulware:1972yco, Boulware:1972zf}. Nevertheless, ghost-free bimetric theory \cite{Hassan:2011ea,Hassan:2011zd,Hassan:2018mbl}, a theory of one massless and one massive spin-2 field, demonstrates that consistent nonlinear interactions are possible. It has also been demonstrated to have interesting phenomenological possibilities, such as observationally viable cosmologies \cite{vonStrauss:2011mq,DeFelice:2014nja,Akrami:2015qga,Hogas:2021fmr,Hogas:2021lns}, and the massive spin-2 mode can provide a dark matter candidate \cite{Aoki:2016zgp,Babichev:2016bxi,Marzola:2017lbt}. Bimetric theory can be extended to several dynamical spin-2 fields, giving rise to multi-gravity theories with interactions that involve more than two fields \cite{Hinterbichler:2012cn,Khosravi:2011zi,Molaee:2018brt,deRham:2015cha,Nomura:2012xr,Scargill:2014wya,Hassan:2018mcw}. Around proportional backgrounds, their linearised spectrum contains one massless and $\Ncal-1$ massive spin-2 modes \cite{Hassan:2012wr,Baldacchino:2016jsz,Flinckman:2024zpb}.

General Relativity has a limited catalogue of exact solutions, and analytic solutions usually rely on special assumptions about the spacetime geometry. In multi-gravity, even fewer exact solutions are known, and their properties are less well understood. While perturbative analysis \cite{Baldacchino:2016jsz,Flinckman:2024zpb,Wood:2025ujj,Flinckman:2026fgu} is useful, exact solutions provide a way to study the additional spin-2 fields beyond perturbation theory. Examples of exact solutions include multi-black holes \cite{Ayon-Beato:2015qtt,Wood:2024acv,Wood:2024eol}, multi-gravity cosmology \cite{Tamanini:2013xia,Luben:2016lku,Khosravi:2011zi,Flinckman:2026fgu}, and AdS waves in three and four spacetime dimensions \cite{Ayon-Beato:2026xaz,Garcia-Compean:2026cnq}, including generalised plane-fronted waves with parallel rays (pp-waves) \cite{Garcia-Compean:2026cnq}. These wave geometries describe exact, nonlinear gravitational waves on AdS spacetime. In this article, we extend the construction of exact gravitational-wave solutions to arbitrary spacetime dimension and to ghost-free multi-gravity interactions beyond pairwise couplings, using the framework developed by Hassan and Schmidt-May \cite{Hassan:2018mcw}. 

Metrics of the form now known as pp-waves already appear in Brinkmann's study of conformally related Einstein spaces \cite{Brinkmann:1925}, while AdS-wave geometries were studied by Siklos \cite{Siklos:1985}. Exact AdS waves have subsequently been found in three-dimensional New Massive Gravity \cite{Ayon-Beato:2009cgh} and Zwei-Dreibein Gravity \cite{Bergshoeff:2014eca}, including logarithmic solutions at special parameter values. In higher dimensions, AdS and pp-waves have been studied in quadratic and generic higher-curvature gravity \cite{Gullu:2011sj,Gurses:2014soa}. Exact pp-waves have also been constructed in ghost-free massive gravity \cite{Mohseni:2011vv}, and AdS and pp-waves in ghost-free bimetric theory \cite{Mohseni:2012ug,Ayon-Beato:2018hxz,Garcia-Compean:2024uie}. Here we construct both types of waves in arbitrary spacetime dimension and for any number of interacting fields in a ghost-free multi-gravity theory with genuine multi-field interactions.

AdS and pp-waves are useful because their profiles can obey linear equations while the corresponding metrics solve the full nonlinear field equations \cite{Gullu:2011sj,Ayon-Beato:2018hxz,Ortaggio:2024say}. They therefore allow us to follow massless and massive spin-2 modes into exact solutions without restricting their amplitudes \cite{Ayon-Beato:2018hxz,Ayon-Beato:2026xaz}. For AdS waves, boundary conditions determine which modes belong to the allowed solution space \cite{Brown:1986,Henneaux:2009pw}. The plane-wave subclass of pp-waves provides an exact setting for studying geodesic deviation and gravitational memory \cite{Zhang:2017rno,Chakraborty:2019yxn}. These connections make exact wave geometries useful for relating the spin-2 spectrum and boundary behaviour of a theory to its nonlinear solutions.

\vspace{3mm}
\noindent\textbf{Summary of results:}
We construct exact AdS waves and pp-waves in ghost-free multi-gravity \cite{Hassan:2018mcw} with genuine interactions among $\Ncal$ vielbeins in arbitrary dimension $D\geq3$. Although the metrics solve the full nonlinear field equations, the wave profiles obey coupled linear equations whose mixing matrix has the same eigenvalues as the quadratic Fierz--Pauli mass matrix found in \cite{Flinckman:2024zpb}. The solutions separate into one massless and $\Ncal-1$ massive eigenprofiles, without any restriction on their amplitudes. In contrast to the critical logarithmic solutions found in Zwei-Dreibein and Viel-dreibein gravity \cite{Bergshoeff:2014eca,Ayon-Beato:2026xaz}, coincident Fierz--Pauli masses do not generate logarithmic modes here: with positive kinetic weights, the mixing matrix remains diagonalisable. We obtain explicit radial and transverse solutions and show that the imposed AdS boundary falloff excludes profiles with negative squared eigenmasses, including logarithmic solutions at the Breitenlohner--Freedman threshold \cite{Breitenlohner:1982jf}. Finally, we relate the AdS waves to pp-waves through a local flat limit.

\vspace{3mm}
\noindent\textbf{Outline:} 
In Section~\ref{sec:multi-gravity}, we introduce the framework of ghost-free multi-gravity and review its relevant properties. We focus on the Hassan--Schmidt-May determinant interaction, which includes genuine multi-vielbein couplings beyond pairwise interactions, and present the field equations and covariant constraints. In Section~\ref{sec:prop_sol}, we show that the theory admits proportional Einstein solutions, which we then specialise to AdS. We deform the proportional AdS vielbeins to obtain AdS waves in Section~\ref{sec:AdS_waves}, derive the wave-profile equations and their solutions, and impose boundary conditions. In Section~\ref{sec:pp-waves}, we construct proportional Minkowski solutions and the corresponding multi-gravity pp-waves. In Section~\ref{sec:examples}, we present explicit examples and show that the determinant interaction admits wave configurations unavailable to standard pairwise tree theories with the same number of fields. We conclude with a discussion of the results and possible extensions.

\vspace{3mm}
\noindent\textbf{Conventions:}
Our analysis applies to arbitrary spacetime dimension $D$ and number of vielbeins $\Ncal$. We use Wald's curvature conventions, the mostly-plus Minkowski metric $\eta_{ab}=\mathrm{diag}(-,+,\ldots,+)$, and units in which the speed of light is one. Spacetime indices are denoted by Greek letters $\mu,\nu,\ldots$, while local Lorentz indices are denoted by Latin letters $a,b,\ldots$. Both are subject to the Einstein summation convention. The metrics, vielbeins, and associated parameters are labelled by capital Latin letters $I,J,\ldots$, ranging from $1$ to $\Ncal$; repeated indices are \textit{not} summed. These labels also carry no tensorial structure, so we place them as superscripts or subscripts wherever this causes the least notational clutter.

\section{Multi-gravity}
\label{sec:multi-gravity}

Multi-gravity is formulated in terms of $\Ncal$ vielbeins $e_I^a=e_I^a{}_\mu\,\mathrm{d}x^\mu$,  each with an Einstein--Hilbert term, together with a non-derivative interaction potential $\Vcal$,\footnote{We assume that the interaction structure is connected, so that all vielbeins belong to a single interacting sector.}
\begin{align}
\label{action}
    \Scal = \int \Bigg[ \sum_{I=1}^{\Ncal}\frac{\mI}{(D-2)!}\epsilon_{a_1\ldots a_D}R_I^{a_1a_2}\wedge e_I^{a_3}\wedge\cdots\wedge e_I^{a_D} - \Vcal(e_1,\ldots,e_{\Ncal}) \Bigg].
\end{align}
The curvature and torsion are defined in terms of the spin connection $\omega_I^{ab}= \omega_I^{ab}{}_\mu\,\mathrm{d}x^\mu$ by
\begin{align}
    R_I^{ab} = \mathrm{d}\omega_I^{ab}+\omega_I^a{}_c\wedge\omega_I^{cb}, \qquad 
    T_I^a = \mathrm{d}e_I^a+\omega_I^a{}_b\wedge e_I^b.
\end{align}
Each vielbein defines an associated metric,
\begin{align}
    g^I_{\mu\nu}=e_I^a{}_\mu\eta_{ab}e_I^b{}_\nu.
\end{align}
The action \eqref{action} is invariant under diagonal local Lorentz transformations $L^a_{~b}(x)$ and diagonal diffeomorphisms,
\begin{align}
    e_I^a\mapsto L^a{}_b\,\varphi^*e_I^b, \qquad I=1,\ldots,\Ncal,
\end{align}
where $\varphi:\mathcal M\to\mathcal M$ is a diffeomorphism and the same $L^a{}_b$ and $\varphi$ act on every vielbein.

For a generic non-derivative potential $\Vcal$, the theory propagates one massless and $\Ncal-1$ massive spin-2 fields, together with $\Ncal-1$ additional scalar modes known as Boulware--Deser ghosts \cite{Boulware:1972yco,Boulware:1972zf}. Seeking interactions without these extra pathological degrees of freedom, Hinterbichler and Rosen proposed a class of antisymmetrised vielbein potentials \cite{Hinterbichler:2012cn},
\begin{align}
\label{pot_wedge}
    \Vcal = \frac{2m^D}{D!}\!\!\!\sum_{I_1,\ldots,I_D=1}^{\Ncal}\!\!\beta^{I_1\ldots I_D}\epsilon_{a_1\ldots a_D}e_{I_1}^{a_1}\wedge\cdots\wedge e_{I_D}^{a_D},
\end{align}
where the dimensionless coefficients $\beta^{I_1\ldots I_D}$ are totally symmetric. However, this potential is not ghost-free for general $\beta$-parameters \cite{Afshar:2014dta, deRham:2015cha}. It was argued in \cite{Flinckman:2026non} that the most general ghost-free interactions within this class are assembled into tree structures built from pairwise bimetric \cite{Hassan:2011zd, Hassan:2011ea, Hassan:2018mbl} or Hassan--Schmidt-May interactions \cite{Hassan:2018mcw}. For the latter, the $\beta$-coefficients involving at least two distinct vielbeins factorise,
\begin{align}
\label{rank-1}
    \beta^{I_1\ldots I_D}=\beta^{I_1}\ldots\beta^{I_D},
\end{align}
while the diagonal coefficients $\beta^{I\ldots I}$ are independent. We will take ${\beta^{I_1\ldots I_D}=\beta^{I_1}\ldots\beta^{I_D}}$ for all $I_1,\ldots, I_D$ and account for the diagonal freedom by including separate cosmological constants for the individual vielbeins, so the action \eqref{action} also contains the cosmological terms $-2\sum_{I=1}^{\Ncal}\mI\Lambda_I\det(e_I)\,\mathrm{d}^Dx$.

Given the decomposition \eqref{rank-1}, we introduce the composite vielbein,
\begin{align}
\label{u_def}
    u^a=\sum_{I=1}^{\Ncal}\beta^I e_I^a.    
\end{align}
The interaction potential \eqref{pot_wedge} can be written as a determinant,
\begin{align}
\label{det_int}
    \Vcal=\frac{2m^D}{D!}\epsilon_{a_1\ldots a_D}u^{a_1}\wedge\cdots\wedge u^{a_D}
    =2m^D\det(u)\,\mathrm{d}^Dx.
\end{align}
We emphasise that this potential contains nontrivial interaction terms beyond pairwise bimetric interactions. For example, for $\Ncal=D=3$, the determinant contains the genuine three-field interaction
\begin{align}
    \Vcal \supset 2m^3\beta^1\beta^2\beta^3\epsilon_{abc}e_1^a\wedge e_2^b\wedge e_3^c.
\end{align}
This term is not ghost-free in isolation \cite{Afshar:2014dta,Flinckman:2026non}, but the determinant also contains terms such as 
\begin{align}
    \Vcal \supset 2m^3 \epsilon_{abc}\big[(\beta^1)^2\beta^{2} e_1^a\wedge e_1^b\wedge e_2^c
    +(\beta^2)^2\beta^{3}e_2^a\wedge e_2^b\wedge e_3^c
    +(\beta^3)^2\beta^{1}e_3^a\wedge e_3^b\wedge e_1^c\big],
\end{align}
whose independent pairwise sum forms a ghostly cycle \cite{Nomura:2012xr,Afshar:2014dta,deRham:2015cha}. In the complete determinant interaction \eqref{det_int}, all these terms have fixed relative coefficients \eqref{rank-1} that ensure the existence of the ghost-eliminating constraints \cite{Flinckman:2025bje, Flinckman:2026non}.

Treating the spin connections and vielbeins as independent variables, the variation with respect to $\omega_I^{ab}$ gives $T_I^a=0$, and the vielbein equations read
\begin{align}
    \frac{\mI}{(D-3)!}\epsilon_{a b_1\ldots b_{D-1}}
    \left[R_I^{b_1b_2}-\frac{2\Lambda_I}{(D-1)(D-2)}e_I^{b_1}\wedge e_I^{b_2}\right]
    \wedge e_I^{b_3}\wedge\cdots\wedge e_I^{b_{D-1}}\notag\\
    -\frac{2m^D\beta^I}{(D-1)!}\epsilon_{a b_1\ldots b_{D-1}}
    u^{b_1}\wedge\cdots\wedge u^{b_{D-1}}=0.
\end{align}
In terms of spacetime indices, the equations can be written as
\begin{align}
\label{eom}
    G^I_{\mu \nu} + \Lambda_I g^I_{\mu \nu} + V^I_{\mu \nu}=0,
\end{align}
where $G^I_{\mu\nu}$ is the Einstein tensor of $g^I_{\mu\nu}$ and
\begin{align}
    \label{V_def}
    V^I_{\mu\nu}=\beta^I\frac{m^D}{\mI}\frac{\det(u)}{\det(e_I)}g^I_{\mu\rho}u^\rho{}_a e_I^a{}_\nu
\end{align}
encodes the multi-gravity interactions. Here $u^\mu_{~a}=(u^{-1})^\mu{}_a$ denotes the inverse composite vielbein, satisfying $u^\mu{}_a u^a{}_\nu=\delta^\mu{}_\nu$, which is guaranteed to exist since $\Vcal \sim \det(u) \neq0$. 

We now briefly count the propagating degrees of freedom. After the torsion equations determine the spin connections, the $\Ncal$ vielbeins contain $\Ncal D^2$ components. The antisymmetric parts of the vielbein equations $V^I_{[\mu \nu]}=0$ impose the algebraic equations,
\begin{align}
\label{LC}
    \beta^I e^a_{I[\mu}\eta_{ab}u^b_{\nu]}=0,
\end{align}
called the Lorentz constraints. Since the sum of \eqref{LC} over $I$ vanishes identically,  only $(\Ncal-1)D(D-1)/2$ of these are independent for generic $e^a_I$.

Taking the $I$-compatible covariant derivative of the vielbein equations \eqref{eom} yields the Bianchi constraints,
\begin{align}
\label{bianchi}
    {}^{I}\!\Dcal_\mu (g^{\mu \sigma}_I V^I_{\sigma \nu})=0,
\end{align}
where $^{I}\!\Dcal$ is the covariant derivative compatible with $g^I_{\mu \nu}$ and $e^a_{I \,\mu}$. Diagonal diffeomorphism invariance relates these $\Ncal D$ equations, leaving $D(\Ncal-1)$ independent constraints.

Fixing the common Lorentz frame and accounting for the $D$ diffeomorphism gauge freedoms and their associated Hamiltonian and momentum constraints leaves
\begin{align}
    \tfrac{1}{2}D(D-3)+\tfrac{1}{2}(\Ncal-1)(D+1)(D-2)+\Ncal-1
\end{align}
degrees of freedom. The last $\Ncal-1$ modes are the potential Boulware--Deser ghosts which are present for generic $\beta^{I_1\ldots I_D}$. For the determinant interaction \eqref{det_int}, it has been argued in \cite{Flinckman:2025bje} that $\Ncal-1$ additional ghost-eliminating constraints exist, leaving
\begin{align}
    N_{\mathrm{DoF}}
    =\tfrac{1}{2}D(D-3)
     +\tfrac{1}{2}(\Ncal-1)(D+1)(D-2),
\end{align}
corresponding to one massless and $\Ncal-1$ massive spin-2 fields.

\section{Proportional Einstein solutions}
\label{sec:prop_sol}

Before proceeding with AdS and pp-wave solutions in Hassan--Schmidt-May multi-gravity, we first establish its proportional Einstein sector.

Consider the ansatz,
\begin{align}
\label{eI=ce}
    e_I^a=c_I\bar e^a,
\end{align}
where the $c_I$ are nonzero real constants and $\bar e^a$ is an invertible vielbein that we will subsequently choose to describe AdS spacetime and later Minkowski.

The composite vielbein \eqref{u_def} is proportional to $\bar{e}^a$,
\begin{align}
    u^a=k\bar e^a,    \qquad k=\sum_{I=1}^{\Ncal}\beta^I c_I,
\end{align}
and we require $k\neq0$, so that $u^a$ is invertible and the potential is nonzero. The torsion equations imply
$\omega_I^{ab}=\bar\omega^{ab}$ for every $I$ and the Lorentz and Bianchi constraints are satisfied identically.

Defining
\begin{align}
    d_I=\beta^I\frac{m^Dk^{D-1}}{\mI c_I^{D-3}},
\end{align}
and substituting \eqref{eI=ce} into the vielbein equations \eqref{eom} yields
\begin{align}
\label{prop-eom}
    \overline G_{\mu\nu}+\left(\Lambda_Ic_I^2+d_I\right)\bar g_{\mu\nu}=0, \qquad I=1,\ldots,\Ncal.
\end{align}
Consequently, $\bar g_{\mu\nu}$ must be an Einstein metric, and the coefficient in parentheses must take the same value for every $I$. Parametrising this value as $\Lambda$, we obtain the consistency conditions
\begin{align}
    \label{consistency_c}
    \Lambda=\Lambda_I c_I^2+d_I, \qquad I=1,\ldots,\Ncal,
\end{align}
on the theory parameters. Conversely, once these algebraic conditions are satisfied, any metric obeying $\overline G_{\mu\nu}+\Lambda\bar g_{\mu\nu}=0$ gives a proportional solution of the full multi-gravity equations. The theory therefore contains an exact Einstein sector.

After fixing the overall normalisation of $\bar e^a$, the $\Ncal$ algebraic equations \eqref{consistency_c} determine the allowed proportionality constants $c_I$ and effective cosmological constant. In \cite{Flinckman:2024zpb} it was shown that \eqref{consistency_c} always has solutions for $c_I$ in the $\beta^Ic_I>0$ quadrant. Alternatively, for prescribed proportionality constants and $\Lambda$, we can solve \eqref{consistency_c} by fixing the bare cosmological constants,
\begin{align}
\label{cc_sol}
    \Lambda_I=\frac{\Lambda-d_I}{c_I^2}.
\end{align}
On the branch $\beta^Ic_I>0$, with positive kinetic weights $\mI c_I^{D-2}>0$, the consistency conditions imply
\begin{align}
    \Lambda-\Lambda_Ic_I^2=d_I>0.
\end{align}
The interaction therefore contributes positively to the effective cosmological constant, without fixing its sign. We now focus on proportional AdS solutions, for which $\Lambda<0$ and hence $\Lambda_I<0$ for every $I$.

\section{Multi AdS waves}
\label{sec:AdS_waves}

Starting from the proportional AdS solutions, we now construct exact gravitational-wave solutions by deforming the vielbeins along a common null direction. The resulting metrics take the Siklos form and correspond to AdS waves.

\subsection{Proportional AdS}

For an AdS spacetime with radius $\ell$, the Einstein tensor is proportional to the metric,
\begin{align}
\label{G_AdS}
    \bar G_{\mu\nu}=\frac{(D-1)(D-2)}{2\ell^2}\bar g_{\mu\nu}, \qquad \Lambda=-\frac{(D-1)(D-2)}{2\ell^2}.
\end{align}
The individual metrics $g^I_{\mu\nu}=c_I^2\bar g_{\mu\nu}$ share the same null cones and isometries, although their physical AdS radii and effective cosmological constants differ,
\begin{align}
    \ell_I=|c_I|\ell, \qquad \Lambda_{\mathrm{eff}}^I=\frac{\Lambda}{c_I^2}.
\end{align}
We now choose the vielbein $\bar e^a$ to represent AdS in the coordinates $x^\mu=(x^+,x^-,z,x^A)$, where ${x^\pm=(t\pm x)/\sqrt{2}}$ are advanced and retarded null coordinates, $z>0$ is the radial coordinate, $x^+,x^-,x^A\in\mathbb R$, and $A=3,\ldots,D-1$. A convenient AdS vielbein is
\begin{equation}
\label{AdS_vielbein}
\begin{aligned}
    \bar e^0 &= \frac{\ell}{\sqrt{2}z}\left[\mathrm{d}x^++\mathrm{d}x^-\right],
    &\quad \bar e^1 &= \frac{\ell}{\sqrt{2}z}\left[\mathrm{d}x^+-\mathrm{d}x^-\right],\\
    \bar e^2 &= \frac{\ell}{z}\,\mathrm{d}z,
    &\quad \bar e^{A} &= \frac{\ell}{z}\,\mathrm{d}x^A.
\end{aligned}
\end{equation}
This yields the AdS metric in the Poincaré patch, with the boundary directions written in light-cone form,
\begin{align}
    \mathrm{d}\bar s^2=\bar g_{\mu \nu} \mathrm{d}x^\mu\mathrm{d}x^\nu=\frac{\ell^2}{z^2}\left[-2\,\mathrm{d}x^+\mathrm{d}x^-+\mathrm{d}z^2+\delta_{AB}\mathrm{d}x^A\mathrm{d}x^B\right],
\end{align}
with the conformal boundary at $z=0$. Substituting this metric into \eqref{prop-eom} with the choice \eqref{cc_sol} clearly yields a proportional AdS branch of multi-gravity.

\subsection{Exact AdS waves}

Starting from \eqref{AdS_vielbein}, we introduce AdS waves by deforming the vielbeins,
\begin{equation}
\label{AdSw_vielein}
\begin{aligned}
    e_I^0 &= c_I\left[\bar e^0+\tfrac{\ell}{2\sqrt{2}z}F_I\,\mathrm{d}x^+\right], & e_I^1 &= c_I\left[\bar e^1-\tfrac{\ell}{2\sqrt{2}z}F_I\,\mathrm{d}x^+\right] \\
    e_I^2 &= c_I\bar e^2, &\qquad 
    e_I^A &= c_I\bar e^A,
\end{aligned}
\end{equation}
where the profiles $F_I=F_I(x^+,z,x^A)$ are independent of $x^-$.\footnote{Allowing $x^-$ dependence does not give additional local solutions within this ansatz. The Bianchi constraints require $\partial_-(F_I-F_J)=0$ for all $I, J$, while the remaining field equations restrict the shared dependence to $A(x^+)x^-$. A common coordinate transformation $x^{+\prime}=h(x^+)$, $x^{-\prime}=x^-/h'(x^+)$ removes this term when $2h''+Ah'=0$.} The associated metrics take the Siklos form,
\begin{align}
\label{AdS_metric}
    \mathrm{d}s_I^2=\frac{c_I^2\ell^2}{z^2}\left[-2\,\mathrm{d}x^+\mathrm{d}x^- -F_I(\mathrm{d}x^+)^2+\mathrm{d}z^2+\delta_{AB}\mathrm{d}x^A\mathrm{d}x^B\right].
\end{align}
Note that the vielbein deformation $ c_I\ell F_I/2\sqrt{2}z$ need not be small, and we will see that the ansatz describes finite-amplitude waves without a perturbative expansion around the proportional AdS solutions. The deformation also leaves the determinant unchanged, $\det(e_I)=c_I^D\det(\bar e)$, so the vielbeins remain invertible and well defined for all finite profiles, while setting the profiles to zero recovers the proportional AdS solutions.

The composite vielbein $u^a$ has the same form as $e_I^a$, with $c_I$ and $F_I$ replaced by
\begin{align}
    k=\sum_{I=1}^{\Ncal}\beta^Ic_I, \qquad F_u=\frac{1}{k}\sum_{I=1}^{\Ncal}\beta^Ic_IF_I.
\end{align}
Let $\overline{\Box}$ denote the scalar d'Alembertian of the common AdS metric $\bar g_{\mu\nu}$. In our coordinates,
\begin{align}
\label{Box}
    \overline{\Box}\psi=\frac{1}{\sqrt{|\bar g|}}\partial_\mu\left(\sqrt{|\bar g|}\,\bar g^{\mu\nu}\partial_\nu\psi\right)=\frac{1}{\ell^2}\left[z^2\left(\partial_z^2+\delta^{AB}\partial_A\partial_B-2\partial_+\partial_-\right)-(D-2)z\partial_z\right]\psi.
\end{align}
Note that the term containing $\partial_-$ vanishes when this operator acts on any of the profiles $F_I$ as they are independent of $x^-$.

With the Siklos metric \eqref{AdS_metric}, the Einstein tensor can be shown to yield the standard AdS term \eqref{G_AdS} with an additional $\overline{\Box}F_I$ contribution,
\begin{align}
\label{G_AdSw}
    G^I_{\mu\nu}=\frac{(D-1)(D-2)}{2c_I^2\ell^2}g^I_{\mu\nu}+\frac{\ell^2}{2z^2}\overline{\Box} F_I\,\delta^+_\mu\delta^+_\nu,
\end{align}
while the multi-gravity interaction term \eqref{V_def} is linear in the wave profiles,
\begin{align}
\label{AdSw-pot}
    V^I_{\mu\nu}=\frac{d_I}{c_I^2}\left[g^I_{\mu\nu}-\frac{c_I^2\ell^2}{2z^2}(F_I-F_u)\delta^+_\mu\delta^+_\nu\right].
\end{align}
This is manifestly symmetric, which makes the Lorentz constraints \eqref{LC}, $V^I_{[\mu \nu]}=0$, trivial while the Bianchi constraints \eqref{bianchi} vanish identically since all profiles are independent of $x^-$.

Substituting \eqref{G_AdSw} and \eqref{AdSw-pot} into the field equations \eqref{eom}, we find that the terms proportional to $g^I_{\mu\nu}$ cancel by the proportional-AdS conditions,
\begin{align}
\label{consistency_AdS}
    \Lambda_Ic_I^2+d_I=-\frac{(D-1)(D-2)}{2\ell^2},
\end{align}
and the remaining field equations yield wave equations for the profiles $F_I$,
\begin{align}
\label{D_wave_eq}
     \overline{\Box} F_I-\sum_{J=1}^{\Ncal}\Mcal_{IJ}F_J=0,
\end{align}
where the profile mixing matrix $\Mcal_{IJ}$ is given by
\begin{align}
    \qquad \Mcal_{IJ}=d_I\left(\delta_{IJ}-\frac{\beta^Jc_J}{k}\right).
\end{align}
It is easy to verify that the rows of $\Mcal$ sum to zero, so that the vector
\begin{align}
    v^I_0=1
\end{align}
is an eigenvector with $\mu_0^2=0$. The remaining eigenvalues are given by the characteristic equation,
\begin{align}
\label{char_eq}
    \det\left(\mu^2\id-\Mcal\right)=\frac{\mu^2}{k}\sum_{I=1}^{\Ncal}\beta^Ic_I\prod_{J\ne I}\left(\mu^2-d_J\right)=0.
\end{align}
The characteristic equation also shows that the zero eigenvalue cannot become degenerate by tuning finite, nonzero interaction coefficients. Comparing the terms linear in $\mu^2$ gives
\begin{align}
    \prod_{n=1}^{\Ncal-1}\mu_n^2
    &=\frac{1}{k}\sum_{I=1}^{\Ncal}\beta^Ic_I
      \prod_{J\ne I}d_J
      =\frac{\sum_{I=1}^{\Ncal}m_I^{D-2}c_I^{D-2}}
             {m^Dk^D}\prod_{I=1}^{\Ncal}d_I.
\end{align}
For finite positive kinetic weights $m_I^{D-2}c_I^{D-2}>0$, ensuring non-ghostly spin-2 kinetic terms, nonzero $m$ and $k$, and $\beta^Ic_I\ne0$ for every $I$, the right-hand side is nonzero. Hence $v_0^I$ spans the unique massless eigenspace. An additional zero eigenvalue at finite regular parameters requires a vanishing interaction coefficient $\beta^I$, in which case the corresponding vielbein decouples.

In \cite{Flinckman:2024zpb}, the squared Fierz--Pauli masses of the quadratic theory expanded around a proportional Einstein solution were shown to obey precisely the characteristic equation \eqref{char_eq}. Thus the eigenvalues $\mu_n^2$ are the squared Fierz--Pauli masses of the associated spin-2 fields. It was also shown that the consistency conditions \eqref{consistency_c} always admit solutions with $\beta^Ic_I>0$, for which the massive eigenvalues are positive. Ordering the $I$-labels so that $d_1\leq\cdots\leq d_{\Ncal}$, the eigenvalues obey \cite{Flinckman:2024zpb}
\begin{align}
\label{sandwitch}
    0=\mu_0^2<d_1\leq\mu_1^2\leq d_2\leq\cdots\leq d_{\Ncal-1}\leq\mu_{\Ncal-1}^2\leq d_{\Ncal}.
\end{align}
The mixing matrix $\Mcal$ is similar to a real symmetric matrix, so its eigenvectors form a complete basis, even when some massive eigenvalues coincide.\footnote{For positive kinetic weights $w_I=m_I^{D-2}c_I^{D-2}>0$, one has $w_I\Mcal_{IJ}=w_J\Mcal_{JI}$ (no sum). Thus $W^{1/2}\Mcal W^{-1/2}$ is real symmetric, where $W=\operatorname{diag}(w_1,\ldots,w_{\Ncal})$.} Although the roots of \eqref{char_eq} are difficult to obtain explicitly for arbitrary $\Ncal$, an eigenvector whose eigenvalue $\mu_n^2$ differs from every $d_I$ takes the form\footnote{If $\mu_n^2=d$ coincides with one or more $d_I$, the normalisation used here is unavailable. The corresponding eigenvectors have $v_n^I=0$ whenever $d_I\ne d$ and satisfy $\sum_{I:\,d_I=d}\beta^Ic_Iv_n^I=0$, where the sum runs only over indices $I$ for which $d_I=d$. These eigenvectors can be obtained directly from the eigenvalue equation. With positive kinetic weights, $\Mcal$ remains diagonalisable also at such degeneracies.}
\begin{align}
\label{eigen_v}
    v^I_{n}=\frac{d_I}{d_I-\mu_n^2},
\end{align}
where we have chosen the normalisation $\sum_J\beta^Jc_Jv^J_{n}/k=1$. Together with $v^I_{0}=1$, these eigenvectors obey
\begin{align}
    \sum_{J=1}^{\Ncal}\Mcal_{IJ}v^J_{n}=\mu_n^2v^I_{n}, \qquad n=0,\ldots,\Ncal-1.
\end{align}
Expanding the profiles as $F_I=\sum_{n=0}^{\Ncal-1}v^I_{n}f_n$, we obtain $\Ncal$ decoupled equations,
\begin{align}
    \left(\overline{\Box}-\mu_n^2\right)f_n=0, \qquad n=0,\ldots,\Ncal-1.
\end{align}
These are Klein--Gordon equations restricted to profiles independent of $x^-$. The dependence on $x^+$ remains arbitrary, while the radial and transverse dependence is determined by $\mu_n^2$. Although the equations are linear, their solutions give exact Siklos spacetimes at finite amplitude, and the modes are deformed by the corresponding Fierz--Pauli masses.

\subsubsection{\texorpdfstring{$D=3$}{D=3}}

In $2+1$ dimensions, the index set $A$ is empty, so the scalar AdS wave operator $\overline{\Box}$ \eqref{Box} simplifies,
\begin{align}
    \overline{\Box}=\frac{1}{\ell^2}\left(z^2\partial_z^2-z\partial_z\right),
\end{align}
and each eigenprofile $f_n$ therefore obeys
\begin{align}
    \left(z^2\partial_z^2-z\partial_z-\ell^2\mu_n^2\right)f_n=0.
\end{align}
From the Frobenius ansatz $f_n\propto(z/\ell)^p$, the indicial equation gives
\begin{align}
    p(p-2)-\ell^2\mu_n^2=0,
\end{align}
which yields the powers,
\begin{align}
    p_n^\pm=1\pm\nu_n, \qquad \nu_n=\sqrt{1+\ell^2\mu_n^2}.
\end{align}
The massless mode $f_0$ has radial powers $0$ and $2$ and is common to all $F_I$. Provided $\nu_n\ne0$ for the massive modes, the general solution is
\begin{align}
\label{D3_gen_sol}
    F_I=a_0(x^+)+a_2(x^+)\left(\frac{z}{\ell}\right)^2+\sum_{n=1}^{\Ncal-1}v^I_n\left[a_n^+(x^+)\left(\frac{z}{\ell}\right)^{1+\nu_n}+a_n^-(x^+)\left(\frac{z}{\ell}\right)^{1-\nu_n}\right].
\end{align}
The $2\Ncal$ coefficient functions $a_0, a_2, a^\pm_n$ are arbitrary functions of $x^+$, subject to the reality of the profiles.

The coefficients $a_0$ and $a_2$ are common to all metrics because both radial powers belong to the massless eigenvector, $v^I_{0}=1$. In $D=3$, both terms can be removed locally by a single coordinate transformation: an $x^-$ shift removes $a_0$, while an $x^+$ reparametrisation removes $a_2$. This local equivalence need not extend to the boundary or respect global identifications, where the massless sector may carry nontrivial data.

By contrast, the massive profiles produce local curvature beyond that of AdS. In three dimensions, the Ricci tensor satisfies
\begin{align}
    R^I_{\mu\nu}+\frac{2}{c_I^2\ell^2}g^I_{\mu\nu}=\frac{\ell^2}{2z^2}\sum_{n=1}^{\Ncal-1}\mu_n^2v^I_n f_n\,\delta^+_\mu\delta^+_\nu.
\end{align}
Since all $\mu_n^2$ in the sum are nonzero and the eigenvectors are linearly independent, the right-hand side vanishes for every $I$ only if all massive profiles vanish. Thus the metrics are all locally AdS only in the absence of massive profiles, which therefore cannot be removed by a common coordinate transformation.

To select solutions with the standard asymptotically AdS$_3$ behaviour, we impose the Brown--Henneaux falloff \cite{Brown:1986, Henneaux:2009pw}. Within our ansatz, its nontrivial condition is
\begin{align}
\label{brown-henneaux}
    g^I_{++}=\mathcal O(1) \qquad \Longleftrightarrow \qquad F_I=\mathcal O(z^2), \qquad z\to0,
\end{align}
which for positive masses requires
\begin{align}
    a_0(x^+)=0, \qquad a^-_n(x^+)=0,
\end{align}
leaving
\begin{align}
\label{3D_sol}
    F_I = a_2(x^+)\left(\frac{z}{\ell}\right)^2+\sum_{n=1}^{\Ncal-1}v^I_na_n^+(x^+)\left(\frac{z}{\ell}\right)^{1+\nu_n}.
\end{align}
We explicitly retain $a_2$ as it might represent nontrivial boundary data.

The non-negativity of the eigenmasses relied on the branch $\beta^Ic_I>0$. Other proportional-AdS solutions can have $\mu_n^2<0$ even with positive kinetic weights $m_Ic_I>0$, so that $\Mcal$ remains diagonalisable with real eigenvalues. For example, when $\Ncal=2$, the nonzero squared eigenmass is
\begin{align}
    \mu^2=m^3k\beta^1\beta^2\left(\frac{c_2}{m_1}+\frac{c_1}{m_2}\right),
\end{align}
which can be negative when the two $\beta^I$ have opposite signs.\footnote{In units with $m=m_1=m_2=c_1=c_2=\ell=1$, take $\beta^1=1$, $\beta^2=-1/2$, $\Lambda_1=-5/4$, and $\Lambda_2=-7/8$. Then $k=1/2$, both proportional-AdS conditions \eqref{consistency_AdS} hold, and the eigenvalues are $\mu_0^2=0$ and $\mu_1^2=-1/2$. Thus $\ell^2\mu_1^2=-1/2>-1$, above the Breitenlohner--Freedman threshold of the scalar profile equation.}

For $-1<\ell^2\mu_n^2<0$, $0<\nu_n<1$ and the powers in \eqref{D3_gen_sol} remain real. At the Breitenlohner--Freedman threshold $\ell^2\mu_n^2=-1$, the two powers coincide and the corresponding eigenprofile becomes
\begin{align}
\label{3D_log}
    f_n=\frac{z}{\ell}\left[b_n(x^+)+\tilde b_n(x^+)\log\left(\frac{z}{\ell}\right)\right].
\end{align}
The logarithm arises from the repeated radial root. For $\ell^2\mu_n^2<-1$, the roots are complex, but a real eigenprofile can be written as
\begin{align}
\label{complex_f}
    f_n=\frac{z}{\ell}\left[b_n(x^+)\cos\left(\sigma_n\log\frac{z}{\ell}\right)+\tilde b_n(x^+)\sin\left(\sigma_n\log\frac{z}{\ell}\right)\right], \qquad \sigma_n=\sqrt{-1-\ell^2\mu_n^2}.
\end{align}
However, none of the profiles (\ref{3D_log}--\ref{complex_f}) with $\mu_n^2<0$ obeys the Brown--Henneaux falloff $F_I=\mathcal{O}(z^2)$. This includes the nonlogarithmic solution obtained by setting $\tilde b_n=0$ in \eqref{3D_log}. Negative squared Fierz--Pauli masses also violate the usual spin-2 unitarity bound in AdS$_3$ \cite{Bergshoeff:2014eca,Bergshoeff:2010iy}, and those below the Breitenlohner--Freedman threshold additionally signal an instability \cite{Breitenlohner:1982jf}. For these reasons, we do not consider these profiles further, even under relaxed boundary conditions.

\subsubsection{\texorpdfstring{$D>3$}{D>3}}

The diagonalisation above also applies for $D>3$, but each eigenprofile $f_n$ now obeys
\begin{align}
\label{D_eigenprofile_eq}
    \left[z^2\left(\partial_z^2+\delta^{AB}\partial_A\partial_B\right)-(D-2)z\partial_z-\ell^2\mu_n^2\right]f_n=0,
\end{align}
which contains no $\partial_+$ derivatives, so its solutions may contain arbitrary functions of $x^+$.

We first consider profiles independent of $x^A$, which generalise the $D=3$ solutions \eqref{3D_sol}. Setting $f_n\propto(z/\ell)^p$ gives
\begin{align}
    p(p-D+1)=\ell^2\mu_n^2, \quad \Rightarrow \quad p_n^\pm=\tfrac{D-1}{2}\pm\nu_n, \qquad \nu_n=\sqrt{\tfrac{(D-1)^2}{4}+\ell^2\mu_n^2}.
\end{align}
For $\nu_n\ne0$, the general solution is
\begin{align}
\label{D_x_indep}
    f_n=a_n^+(x^+)\left(\frac{z}{\ell}\right)^{p_n^+}+a_n^-(x^+)\left(\frac{z}{\ell}\right)^{p_n^-}.
\end{align}
In the massless sector ($\mu_0^2=0$) the powers are $0$ and $D-1$. Hence
\begin{align}
\label{mu2=0}
    f_0=a_0(x^+)+a_{D-1}(x^+)\left(\frac{z}{\ell}\right)^{D-1}.
\end{align}
Unlike in $D=3$, the latter solution generally has nonzero Weyl curvature and cannot be removed by a local diffeomorphism. 
At $\ell^2\mu_n^2=-(D-1)^2/4$, the roots coincide ($\nu_n=0$) and the corresponding solution instead reads
\begin{align}
\label{D_x_indep_overlapp}
    f_n=\left(\frac{z}{\ell}\right)^{(D-1)/2}\left[b_n(x^+)+\tilde b_n(x^+)\log\left(\frac{z}{\ell}\right)\right],
\end{align}
generalising the logarithmic solution \eqref{3D_log}.

Each radial solution above may also be multiplied by a function harmonic in the additional transverse coordinates $x^A$. For example,
\begin{align}
\label{harmonic_dep}
    f_n=\left(\frac{z}{\ell}\right)^{p_n^\pm}H_n(x^+,x^A), \qquad \delta^{AB}\partial_A\partial_BH_n=0.
\end{align}
The same construction applies to the logarithmic radial solution \eqref{D_x_indep_overlapp}. 

To obtain more general transverse dependence, we Fourier transform in $\mathbf{x}=(x^A)$,
\begin{align}
    f_n(x^+,z,\mathbf{x})=\!\int\!\frac{\mathrm{d}^{D-3}\mathbf{q}}{(2\pi)^{D-3}}\,e^{i\mathbf{q}\cdot\mathbf{x}}\widehat f_n(x^+,z,\mathbf{q}), \qquad q=|\mathbf{q}|>0.
\end{align}
For each transverse momentum, \eqref{D_eigenprofile_eq} becomes
\begin{align}
    z^2\partial_z^2\widehat f_n-(D-2)z\partial_z\widehat f_n-\left(q^2z^2+\ell^2\mu_n^2\right)\widehat f_n=0.
\end{align}
The substitution $\widehat f_n=(z/\ell)^{(D-1)/2}\phi_n(qz)$ yields the modified Bessel equation,
\begin{align}
    t^2\phi_n''+t\phi_n'-\left(t^2+\nu_n^2\right)\phi_n=0, \qquad t=qz.
\end{align}
Its standard solutions are the modified Bessel functions of the first and second kind, $I_\nu$ and $K_\nu$, which may be written as \cite{ArfkenWeberHarris2013}
\begin{align}
    I_\nu(t)&=\sum_{s=0}^{\infty}\frac{1}{s!\,\Gamma(s+\nu+1)}\left(\frac{t}{2}\right)^{2s+\nu},\\
    K_\nu(t)&=\frac{\pi}{2\sin(\pi\nu)}\left[I_{-\nu}(t)-I_\nu(t)\right], \qquad \nu\notin\mathbb{Z},
\end{align}
where $K_\nu$ at integer order is defined by continuity. Consequently, the general radial solution at nonzero transverse momentum is
\begin{align}
\label{D_Bessel_solution}
    \widehat f_n=\left(\frac{z}{\ell}\right)^{(D-1)/2}\left[A_n(x^+,\mathbf{q})I_{\nu_n}(qz)+B_n(x^+,\mathbf{q})K_{\nu_n}(qz)\right], \qquad q>0.
\end{align}
The coefficients $A_n$ and $B_n$ must be chosen so that the reconstructed profiles are real and satisfy any imposed boundary conditions. The zero-momentum ($q=0$) equation is the Euler equation with solutions \eqref{D_x_indep} and \eqref{D_x_indep_overlapp}. These can also be recovered as $q\to0$ limits of \eqref{D_Bessel_solution} with suitably $q$-dependent coefficients.\footnote{In particular, when $\nu_n=0$, the $\log q$ term in the small-argument expansion of $K_0(qz)$ can be absorbed into the coefficient of $I_0(qz)$, leaving the logarithm in \eqref{D_x_indep_overlapp}. For real $\nu_n>0$, the $I_{\nu_n}$ and $K_{\nu_n}$ branches have leading near-boundary powers $z^{p_n^+}$ and $z^{p_n^-}$, respectively; their behaviour at large $z$ raises a separate question of interior regularity.}

Finally, Fourier transforming back and reconstructing
\begin{align}
    F_I=\sum_{n=0}^{\Ncal-1}v^I_{n}f_n
\end{align}
gives exact solutions of the coupled field equations, without restricting their amplitudes.

A natural AdS falloff for the metrics in $D$ dimensions is given by
\begin{align}
\label{falloff}
    g^I_{++}=\mathcal{O}(z^{D-3}), \qquad z\to0.
\end{align}
This reduces precisely to the Brown--Henneaux condition \eqref{brown-henneaux} in $D=3$ and amounts to the falloff
\begin{align}
    F_I=\mathcal{O}(z^{D-1}),
\end{align}
for the profiles at the conformal boundary.

Since the eigenvectors $v^I_{n}$ form a $z$-independent basis, imposing the falloff on every metric is equivalent to imposing it on every eigenprofile,
\begin{align}
    F_I=\mathcal O(z^{D-1})
    \qquad\Longleftrightarrow\qquad
    f_n=\mathcal O(z^{D-1}).
\end{align}
For solutions with harmonic transverse dependence \eqref{harmonic_dep}, this selects
\begin{align}
    \mu_0^2=0 &: \qquad a_0=0,\qquad
    f_0=a_{D-1}(x^+)\left(\frac{z}{\ell}\right)^{D-1}H_0(x^+,x^A),\\
    \mu_n^2>0 &: \qquad a_n^-=0,\qquad
    f_n=a_n^+(x^+)\left(\frac{z}{\ell}\right)^{p_n^+}H_n(x^+,x^A),
    \qquad p_n^+>D-1.
\end{align}
If the squared Fierz--Pauli masses are negative but above the Breitenlohner--Freedman bound, i.e. ${-(D-1)^2/(4\ell^2)<\mu_n^2<0}$, both powers lie between $0$ and $D-1$, so neither term $z^{p^\pm_n}$ in \eqref{D_x_indep} has the required falloff. At the Breitenlohner--Freedman threshold, both solutions have the power $z^{(D-1)/2}$, one multiplied by a logarithm; below it, the solutions oscillate in $\log z$ with the same overall power. Thus, $f_n=0$ for every mode with a negative squared eigenmass $\mu_n^2<0$.

For $q>0$ and $\nu_n>0$, the Fourier solutions \eqref{D_Bessel_solution} behave near the boundary as
\begin{align}
    \left(\frac{z}{\ell}\right)^{(D-1)/2}I_{\nu_n}(qz)\sim z^{p_n^+}, \qquad
    \left(\frac{z}{\ell}\right)^{(D-1)/2}K_{\nu_n}(qz)\sim z^{p_n^-}.
\end{align}
The required falloff therefore sets $B_n=0$ for $\mu_n^2\geq0$ and excludes both solutions for $\mu_n^2<0$. The allowed nonzero-momentum profiles are
\begin{align}
\label{D_Fourier_boundary_solution}
    f_n(x^+,z,\mathbf{x})
    =\left(\frac{z}{\ell}\right)^{(D-1)/2}
    \int_{\mathbf{q}\ne0}\frac{\mathrm{d}^{D-3}\mathbf{q}}{(2\pi)^{D-3}}\,
    e^{i\mathbf{q}\cdot\mathbf{x}}A_n(x^+,\mathbf{q})I_{\nu_n}(qz),
    \qquad \mu_n^2\geq0.
\end{align}
The zero-momentum solutions are given separately by the allowed power-law terms above.

For the reconstructed profiles $F_I=\sum_n v^I_n f_n$, the imposed falloff retains the common massless mode and the positive-mass eigenmodes, with near-boundary powers $D-1$ and $p_n^+$, respectively, while excluding every nonzero profile associated with a negative eigenvalue. On the branch $\beta^Ic_I>0$ of \eqref{consistency_c}, all eigenvalues are non-negative, so the full set of $\Ncal$ eigenmodes is compatible with this boundary condition.

The falloff \eqref{falloff} constrains the profiles at the conformal boundary, $z=0$, but does not establish regularity at the Poincar\'e horizon, $z\to\infty$. We leave this separate question of horizon regularity for future analysis.

\section{Multi pp-waves}
\label{sec:pp-waves}

We now construct exact wave solutions around proportional Minkowski metrics. The construction follows the same steps as for the AdS waves: we first determine the proportional solutions, deform the vielbeins along a common null direction, and diagonalise the resulting profile equations.

\subsection{Proportional Minkowski solutions}

We use coordinates $x^\mu=(x^+,x^-,y,x^A)$, where $x^\pm=(t\pm x)/\sqrt{2}$ and $A=3,\ldots,D-1$. All coordinates range over $\mathbb R$, and the index set $A$ is empty for $D=3$. A convenient Minkowski vielbein is
\begin{equation}
\label{Minkowski_vielbein}
\begin{aligned}
    \tilde e^0 &= \tfrac{1}{\sqrt{2}}\left[\mathrm dx^++\mathrm dx^-\right], & \qquad \tilde e^1 &= \tfrac{1}{\sqrt{2}}\left[\mathrm dx^+-\mathrm dx^-\right],\\
    \tilde e^2 &= \mathrm dy, & \tilde e^A &= \mathrm dx^A,
\end{aligned}
\end{equation}
with associated metric
\begin{align}
    \mathrm d\tilde s^2=\tilde g_{\mu\nu}\mathrm dx^\mu\mathrm dx^\nu=-2\,\mathrm dx^+\mathrm dx^-+\mathrm dy^2+\delta_{AB}\mathrm dx^A\mathrm dx^B.
\end{align}
The proportional vielbein ansatz \eqref{eI=ce} gives
\begin{align}
    e_I^a=c_I\tilde e^a,\qquad u^a=k\tilde e^a,\qquad k=\sum_{I=1}^{\Ncal}\beta^Ic_I,
\end{align}
where $c_I\ne0$ and $k\ne0$ ensure that the individual and composite vielbeins are invertible.

Since $\widetilde G_{\mu\nu}=0$, the proportional-background equations \eqref{prop-eom} reduce to
\begin{align}
\label{Minkowski_consistency}
    \Lambda_Ic_I^2+d_I=0,\qquad d_I=\frac{m^D\beta^Ik^{D-1}}{\mI c_I^{D-3}},\qquad I=1,\ldots,\Ncal.
\end{align}
For prescribed proportionality constants and interaction parameters, these conditions can be satisfied by choosing $\Lambda_I=-d_I/c_I^2$. On the branch $\beta^Ic_I>0$, with positive kinetic weights $\mI c_I^{D-2}>0$, the interaction contributions $d_I>0$ are therefore cancelled by negative bare cosmological constants.

\subsection{Exact pp-wave profiles}

Starting from the proportional Minkowski solutions, we deform the vielbeins \eqref{Minkowski_vielbein},
\begin{equation}
\label{pp_vielbein}
\begin{aligned}
    e_I^0 &= c_I\left[\tilde e^0+\tfrac{1}{2\sqrt{2}}F_I\mathrm dx^+\right], &e_I^1 &= c_I\left[\tilde e^1-\tfrac{1}{2\sqrt{2}}F_I\mathrm dx^+\right],\\
     e_I^2 &= c_I\tilde e^2, & \qquad e_I^A &= c_I\tilde e^A,
\end{aligned}
\end{equation}
where $F_I=F_I(x^+,y,x^A)$ are independent of $x^-$. The associated metrics are
\begin{align}
\label{pp_metric}
    \mathrm ds_I^2=c_I^2\left[-2\,\mathrm dx^+\mathrm dx^--F_I(\mathrm dx^+)^2+\mathrm dy^2+\delta_{AB}\mathrm dx^A\mathrm dx^B\right].
\end{align}
The vector $\partial_-$ is null and covariantly constant with respect to every metric, so these geometries are pp-waves \cite{Brinkmann:1925}. Setting all profiles to zero recovers the proportional Minkowski background.

The composite vielbein has the same form, with $c_I$ and $F_I$ replaced by
\begin{align}
    k=\sum_{I=1}^{\Ncal}\beta^Ic_I,\qquad F_u=\frac{1}{k}\sum_{I=1}^{\Ncal}\beta^Ic_IF_I.
\end{align}
As for the AdS waves, the deformation preserves the determinants,
\begin{align}
    \det(e_I)=c_I^D\det(\tilde e),\qquad \det(u)=k^D\det(\tilde e),
\end{align}
so the vielbeins remain regular for all finite profiles. 

The potential takes a similar form to the AdS wave expression \eqref{AdSw-pot}
\begin{align}
\label{Minkowski_V}
    V^I_{\mu\nu}=\frac{d_I}{c_I^2}\left[g^I_{\mu\nu}-\frac{c_I^2}{2}(F_I-F_u)\delta^+_\mu\delta^+_\nu\right].
\end{align}
This is manifestly symmetric, so the Lorentz constraints \eqref{LC} are satisfied. Its covariant divergence also vanishes: $\mathrm dx^+$ is covariantly constant, and the remaining derivative is proportional to $\partial_-(F_I-F_u)=0$. Thus the Bianchi constraints \eqref{bianchi} impose no additional conditions within this ansatz.

The scalar wave operator of the common Minkowski metric is
\begin{align}
    \widetilde\Box=-2\partial_+\partial_-+\tilde\Delta,\qquad \tilde\Delta=\partial_y^2+\delta^{AB}\partial_A\partial_B.
\end{align}
On the $x^-$-independent profiles, $\widetilde\Box F_I=\tilde\Delta F_I$. The Ricci scalar vanishes, and the Ricci and Einstein tensors have only a $++$ component,
\begin{align}
\label{Minkowski_R}
    R_I=0,\qquad R^I_{\mu\nu}=G^I_{\mu\nu}=\tfrac12\tilde\Delta F_I\,\delta^+_\mu\delta^+_\nu.
\end{align}
Substituting \eqref{Minkowski_V} and \eqref{Minkowski_R} into the equations of motion \eqref{eom} and using the bare cosmological constants \eqref{Minkowski_consistency}, we obtain the coupled profile equations
\begin{align}
\label{pp_profile_eq}
    \tilde\Delta F_I-\sum_{J=1}^{\Ncal}\Mcal_{IJ}F_J=0,\qquad \Mcal_{IJ}=d_I\left(\delta_{IJ}-\frac{\beta^Jc_J}{k}\right).
\end{align}
The mixing matrix has the same algebraic form as for the AdS waves, now evaluated on the chosen Minkowski solution. Its characteristic equation is again \eqref{char_eq}, and its eigenvalues $\mu_n^2$ are the squared Fierz--Pauli masses
\cite{Flinckman:2024zpb}, and the eigenbasis is complete with real eigenvalues. We can therefore expand
\begin{align}
    F_I=\sum_{n=0}^{\Ncal-1}v^I_n f_n,\qquad \sum_{J=1}^{\Ncal}\Mcal_{IJ}v^J_n=\mu_n^2v^I_n,
\end{align}
so the equations decouple into
\begin{align}
\label{pp_eigenprofile_eq}
    \left(\tilde\Delta-\mu_n^2\right)f_n=0,\qquad n=0,\ldots,\Ncal-1.
\end{align}
These are Minkowski Klein--Gordon equations restricted to profiles independent of $x^-$. Their dependence on $x^+$ remains arbitrary, while their transverse dependence is determined by the masses. The resulting metrics solve the full field equations without restricting the profile amplitudes.

\subsubsection{\texorpdfstring{$D=3$}{D=3}}

For $D=3$, there is only one transverse coordinate, and \eqref{pp_eigenprofile_eq} becomes an ordinary differential equation in $y$. On the positive-mass branch, its general solutions are
\begin{align}
    f_0&=a_0(x^+)+b_0(x^+)y,\\
    f_n&=a_n^+(x^+)e^{\mu_n y}+a_n^-(x^+)e^{-\mu_n y},\qquad \mu_n=\sqrt{\mu_n^2}>0,\qquad n=1,\ldots,\Ncal-1.
\end{align}
The coefficient functions are arbitrary functions of $x^+$. The common affine profile $f_0$ can be removed locally by a single coordinate transformation acting on every metric. The massive profiles instead produce Ricci curvature through $\tilde\Delta f_n=\mu_n^2f_n$ and cannot all be removed by a common coordinate transformation.

\subsubsection{\texorpdfstring{$D>3$}{D>3}}

For $D>3$, the massless profile can be any function harmonic in the full transverse coordinates $(y,x^A)$. The corresponding common-profile sector is Ricci flat but can have nonzero Riemann curvature, describing Einstein pp-waves. A simple transverse dependence for the massive profiles is obtained by multiplying their exponential solutions by a harmonic function,
\begin{align}
    f_n=\left[a_n^+(x^+)e^{\mu_n y}+a_n^-(x^+)e^{-\mu_n y}\right]H_n(x^+,x^A),\qquad \delta^{AB}\partial_A\partial_BH_n=0.
\end{align}
More generally, Fourier transforming in $\mathbf x=(x^A)$ gives
\begin{align}
    f_n(x^+,y,\mathbf x)=\int\frac{\mathrm d^{D-3}\mathbf q}{(2\pi)^{D-3}}\,e^{i\mathbf q\cdot\mathbf x}\widehat f_n(x^+,y,\mathbf q).
\end{align}
For each transverse momentum, the equation becomes
\begin{align}
    \partial_y^2\widehat f_n-\left(q^2+\mu_n^2\right)\widehat f_n=0,\qquad q=|\mathbf q|.
\end{align}
On the non-negative-mass branch $\mu_n^2\geq0$, the solutions are
\begin{align}
\label{pp_Fourier_solution}
    \widehat f_n&=A_n(x^+,\mathbf q)+B_n(x^+,\mathbf q)y, \qquad q^2=\mu_n^2=0,\\
    \widehat f_n&=A_n(x^+,\mathbf q)e^{\sqrt{q^2+\mu_n^2}\,y}+B_n(x^+,\mathbf q)e^{-\sqrt{q^2+\mu_n^2}\,y}, \qquad q^2+\mu_n^2>0.
\end{align}
For the inverse Fourier transform $f_n$ to be real, the coefficients must satisfy
\begin{align}
    A_n(x^+,-\mathbf q)=\overline{A}_n(x^+,\mathbf q), \qquad B_n(x^+,-\mathbf q)=\overline{B}_n(x^+,\mathbf q).
\end{align}
Inverse Fourier transformation and reconstruction through $F_I=\sum_n v^I_n f_n$ then give the exact coupled solutions within the Fourier-transformable class.

Note that nonzero massive profiles cannot be both smooth and bounded on the entire transverse space $\mathbb R^{D-2}$.\footnote{Mohseni \cite{Mohseni:2011vv} constructs exact pp-waves in dRGT massive gravity with profile equation $\Delta_\perp F=a_2m^2F$. For $a_2<0$, identified there as non-tachyonic, this equation admits bounded oscillatory profiles. However, in our curvature conventions, the Einstein tensor in their Eq.~(23) has the opposite sign. Expanding their action gives $m_{\mathrm{FP}}^2=-a_2m^2>0$, and the corrected profile equation is therefore $\Delta_\perp F=m_{\mathrm{FP}}^2F$, consistent with the restriction stated here. The apparent difference thus stems from a sign discrepancy in their displayed equations.} Indeed, at each $x^+$, Fourier transformation in all transverse coordinates gives
\begin{align}
    \left(|\mathbf p|^2+\mu_n^2\right)\widetilde f_n(\mathbf p)=0.
\end{align}
For $\mu_n^2>0$, the multiplier never vanishes, so $f_n=0$. The Fourier transform may here be understood in the sense of tempered distributions, which includes bounded profiles. This restriction follows from the assumed independence of $x^-$. We impose no transverse falloff in this construction and leave questions of global regularity for future analysis.

\subsection{Flat limit of the AdS waves}

The pp-waves can also be obtained as a local flat limit of the AdS-wave solutions. Consider a family of proportional AdS solutions with $\ell\to\infty$. We set $z=\ell+y$ and hold $y$, $x^\pm$, and $x^A$ fixed, together with the parameters $c_I,\beta^I,m,m_I$. The bare cosmological constants vary with $\ell$ according to
\begin{align}
    \Lambda_I(\ell)c_I^2+d_I=-\frac{(D-1)(D-2)}{2\ell^2}.
\end{align}
Consequently, $\Lambda_I(\ell)c_I^2+d_I\to0$, as required by the Minkowski conditions \eqref{Minkowski_consistency}, while the mixing matrix and its eigenvalues remain fixed.

On profiles independent of $x^-$, the AdS wave operator becomes
\begin{align}
    \overline{\Box}=\left(1+\frac{y}{\ell}\right)^2\tilde\Delta-\frac{D-2}{\ell}\left(1+\frac{y}{\ell}\right)\partial_y\longrightarrow\tilde\Delta,\qquad \ell\to\infty.
\end{align}
The conformal factor of the AdS metrics similarly satisfies $\ell^2/z^2=(1+y/\ell)^{-2}\to1$. Hence AdS solutions whose profiles have finite limits give the pp-wave metrics \eqref{pp_metric} and their profile equations.

For example, for $\mu_n^2>0$ and no $x^A$ dependence, $p_n^\pm/\ell\to\pm\mu_n$. The two AdS radial solutions therefore become
\begin{align}
    \left(\frac{z}{\ell}\right)^{p_n^\pm}=\left(1+\frac{y}{\ell}\right)^{p_n^\pm}\longrightarrow e^{\pm\mu_n y}.
\end{align}
In the massless sector, a suitable combination of the constant and $z^{D-1}$ solutions gives the linear pp-wave profile,
\begin{align}
    \frac{\ell}{D-1}\left[\left(\frac{z}{\ell}\right)^{D-1}-1\right]\longrightarrow y.
\end{align}
The AdS boundary lies at $y=-\ell$ and recedes to infinity in this limit. Its falloff condition therefore places no boundary condition at finite $y$ in the resulting flat spacetime.

\section{Explicit solutions}
\label{sec:examples}

We now give some examples illustrating regular mass degeneracy, distinct radial falloffs, and the construction of a prescribed mass spectrum. These also clarify the relation to previous wave solutions and provide explicit families on both AdS and Minkowski spacetimes.

\subsection{Degenerate masses in \texorpdfstring{$D=\Ncal=3$}{D=N=3}}

For a simple three-dimensional example with three vielbeins, choose, in units with $\ell=1$,
\begin{align}
    m=c_I=\beta^I=1, \qquad m_I=3, \qquad \Lambda_I=-4, \qquad I=1,2,3.
\end{align}
These parameters satisfy the proportional-AdS conditions \eqref{consistency_c}, with $k=3$ and $d_I=3$, and preserve permutation symmetry among the three vielbeins. The mixing matrix has the spectrum
\begin{align}
    \Mcal=\begin{pmatrix}2&-1&-1\\-1&2&-1\\-1&-1&2\end{pmatrix}, \qquad \mu_0^2=0, \qquad \mu_1^2=\mu_2^2=3.
\end{align}
The massive eigenspace consists of vectors whose components sum to zero, and a basis can be chosen as $(1,0,-1)$ and $(0,1,-1)$. The massive radial powers are $p^+=3$ and $p^-=-1$. Retaining the branches that obey the Brown--Henneaux falloff gives
\begin{align}
    F_1&=a(x^+)z^2+b_1(x^+)z^3,\\
    F_2&=a(x^+)z^2+b_2(x^+)z^3,\\
    F_3&=a(x^+)z^2-\bigl[b_1(x^+)+b_2(x^+)\bigr]z^3.
\end{align}
The three coefficient functions are arbitrary and real. All three metrics may therefore have distinct profiles, while their composite profile contains only the common massless contribution,
\begin{align}
    F_u=\frac{F_1+F_2+F_3}{3}=a(x^+)z^2.
\end{align}
As discussed above, this common term is locally removable in three dimensions, but is retained because it may carry boundary data.

The coincident massive eigenvalues correspond to two independent eigenvectors, each carrying its own massive profile. No logarithmic partners arise because the matrix satisfies
\begin{align}
    \Mcal^2=3\Mcal \qquad\Longrightarrow\qquad \overline{\Box}(\overline{\Box}-3)F_I=0.
\end{align}
Thus each component obeys an equation with two distinct Klein--Gordon factors, despite the repeated root in the characteristic polynomial of $\Mcal$.

This regular degeneracy distinguishes the example from three-field theories with pairwise interactions on a tree \cite{Afshar:2014dta}. Such a tree contains only two interacting pairs, so its profile mixing matrix has a vanishing off-diagonal pair; after relabelling, $(\Mcal_{\mathrm{pair}})_{13}=(\Mcal_{\mathrm{pair}})_{31}=0$. However, with positive kinetic weights, a matrix with the common massless eigenvector and two independent eigenvectors of eigenvalue $3$ would necessarily have the spectral decomposition
\begin{align}
    (\Mcal_{\mathrm{pair}})_{IJ}=3\left(\delta_{IJ}-\frac{m_Jc_J}{\sum_{K=1}^{3}m_Kc_K}\right).
\end{align}
Every off-diagonal entry is then nonzero, requiring an interaction between all three pairs. This is incompatible with a tree, irrespective of the positive kinetic weights or the bimetric potential chosen on each link. Consequently, the full two-function massive sector above cannot be reproduced by a three-field pairwise tree theory.

This also clarifies the comparison with the equal-mass critical solutions of \cite{Ayon-Beato:2026xaz}. At finite, nonzero kinetic weights, their equal-mass logarithmic partners require mixed signs among the coefficients $\sigma_I\gamma_I$ of the quadratic Einstein--Hilbert terms \cite{Afshar:2014dta}. Otherwise, the profile mixing matrix would be similar to a real symmetric matrix and could not develop a Jordan block. For matching squared masses $M_+^2=M_-^2=3$, the repeated eigenvalue has only one independent eigenvector, and their single-profile equation retains the factor $(\overline{\Box}-3)^2$, admitting $z^3\log z$ partners as in their Eqs.~(36) and~(45). The kinetic form on this massive sector has both positive and negative directions, so retaining the logarithmic partners introduces a spin-2 ghost; no real invertible field redefinition can remove this indefinite signature.\footnote{A matching parameter choice in the notation of \cite{Ayon-Beato:2026xaz} is $\ell=m=\gamma_I=1$, $(\sigma_1,\sigma_2,\sigma_3)=(1,1,-1)$, $\beta_{12}=3$, $\beta_{23}=-3$, and $(\alpha_1,\alpha_2,\alpha_3)=(7,-2,-4)$. These parameters satisfy their proportional-AdS conditions and give eigenvalues $0,3,3$, with a one-dimensional eigenspace at the repeated eigenvalue. In the basis $(1,0,1)$ and $(0,-1,-1)$ of its massive generalised eigenspace, the kinetic matrix is proportional to $\left(\begin{smallmatrix}0&1\\1&0\end{smallmatrix}\right)$, explicitly exhibiting opposite kinetic signs.}

\subsection{Distinct masses in \texorpdfstring{$D=4$}{D=4}}

For an example with three vielbeins in four dimensions and distinct massive eigenvalues, choose
\begin{align}
    \ell=m=c_1=c_2=c_3=1, \qquad (\beta^1,\beta^2,\beta^3)=(1,1,2), \qquad (m_1,m_2,m_3)=(4,4,2).
\end{align}
We then have $k=4$ and $(d_1,d_2,d_3)=(4,4,32)$. The proportional-AdS conditions \eqref{consistency_AdS} are satisfied by
\begin{align}
    (\Lambda_1,\Lambda_2,\Lambda_3)=(-7,-7,-35).
\end{align}
The mixing matrix is
\begin{align}
    \Mcal=\begin{pmatrix}3&-1&-2\\-1&3&-2\\-8&-8&16\end{pmatrix}.
\end{align}
The kinetic weights are $(16,16,4)$, and a basis of eigenvectors and their eigenvalues is
\begin{align}
    v_{0}&=(1,1,1),& \mu_0^2&=0,\\
    v_{1}&=(1,-1,0),& \mu_1^2&=4,\\
    v_{2}&=(1,1,-8),& \mu_2^2&=18.
\end{align}
Here $\mu_1^2=d_1=d_2$, so this example realises equality in \eqref{sandwitch}. Its eigenvector must be obtained directly from the eigenvalue equation, rather than from the expression ${d_I/(d_I-\mu_1^2)}$ \eqref{eigen_v}.

The boundary-admissible radial powers are $p_0^+=3$, $p_1^+=4$, and $p_2^+=6$. In the sector without transverse dependence, the solution is
\begin{align}
    F_1&=a(x^+)z^3+b(x^+)z^4+c(x^+)z^6,\\
    F_2&=a(x^+)z^3-b(x^+)z^4+c(x^+)z^6,\\
    F_3&=a(x^+)z^3-8c(x^+)z^6.
\end{align}
The three coefficient functions are arbitrary and real, and all profiles satisfy $F_I=\mathcal O(z^3)$. The common $z^3$ contribution is an Einstein AdS wave with nonzero Weyl curvature when $a\ne0$, so it is not locally removable as in the three-dimensional example.

This example also distinguishes the composite profile from the massless eigenprofile:
\begin{align}
    F_u&=\frac{F_1+F_2+2F_3}{4}=a(x^+)z^3-\frac72c(x^+)z^6,\\
    f_0&=\frac{4F_1+4F_2+F_3}{9}=a(x^+)z^3.
\end{align}
The first massive mode leaves $F_u$ unchanged and does not contribute to $F_3$, while the second contributes to the composite profile. Thus $F_u$ need not coincide with the common massless eigenprofile.

We can also give the two massive modes transverse dependence. Their Bessel orders are $\nu_1=5/2$ and $\nu_2=9/2$, so for any $q>0$ another exact solution is
\begin{align}
    F_1&=a(x^+)z^3+b(x^+)z^{3/2}I_{5/2}(qz)\cos(qx^3)+c(x^+)z^{3/2}I_{9/2}(qz)\cos(qx^3),\\
    F_2&=a(x^+)z^3-b(x^+)z^{3/2}I_{5/2}(qz)\cos(qx^3)+c(x^+)z^{3/2}I_{9/2}(qz)\cos(qx^3),\\
    F_3&=a(x^+)z^3-8c(x^+)z^{3/2}I_{9/2}(qz)\cos(qx^3).
\end{align}
The massive terms begin at orders $z^4$ and $z^6$, respectively, near the conformal boundary. With suitable $q$-dependent rescalings of $b$ and $c$, the limit $q\to0$ recovers the radial solution above. These half-integer-order modified Bessel functions can also be expressed using hyperbolic functions. Their growth at large $z$ means that this example imposes the boundary falloff without an additional regularity condition at the Poincar\'e horizon.

When both massive modes are excited, these profiles cannot be reproduced by a three-field theory with pairwise interactions on a tree. Pairwise profile equations involve differences $F_I-F_J$ only between fields that interact directly. Matching the independent $z^4$ and $z^6$ contributions fixes the first two equations to be
\begin{align}
    \bar\Box F_1&=4bz^4+18cz^6=(F_1-F_2)+2(F_1-F_3),\\
    \bar\Box F_2&=-4bz^4+18cz^6=(F_2-F_1)+2(F_2-F_3).
\end{align}
The first equation requires field $1$ to interact with both fields $2$ and $3$, while the second additionally requires an interaction between fields $2$ and $3$. Reproducing these simultaneous profiles therefore requires three pairwise links, forming a triangle rather than a tree. This obstruction persists when the full bimetric potential is allowed on each link. The same argument applies to the Bessel solutions, whose two massive radial functions are also linearly independent. Thus the determinant interaction admits metric profiles unavailable to the standard three-field pairwise tree theories.

\subsection{Distinct masses for arbitrary \texorpdfstring{$\Ncal$}{N}}

We now give an explicit family of multi-gravity AdS waves with a prescribed nondegenerate mass spectrum in any dimension $D\geq3$. Let $m,\mu>0$ and choose
\begin{align}
    c_I=1,\qquad d_I=\left(I-\tfrac12\right)\mu^2,\qquad \beta^I=\frac{1}{4^{\Ncal-1}}\binom{2I-2}{I-1}\binom{2\Ncal-2I}{\Ncal-I}.
\end{align}
All the $\beta^I$ are positive and $\sum_I\beta^I=1$, so that $k=1$. The chosen values of $d_I$ are realised by
\begin{align}
    m_I^{D-2}=\frac{m^D\beta^I}{d_I}.
\end{align}
Taking
\begin{align}
    \Lambda_I=-\frac{(D-1)(D-2)}{2\ell^2}-d_I
\end{align}
then satisfies the proportional-AdS conditions \eqref{consistency_c} for any $\ell>0$.

To determine the spectrum, observe that the coefficients can equivalently be written as
\begin{align}
    \beta^I=\frac{\prod_{n=1}^{\Ncal-1}(d_I-n\mu^2)}{\prod_{J\ne I}(d_I-d_J)}.
\end{align}
The characteristic polynomial therefore becomes
\begin{align}
    \det\left(\lambda\id-\Mcal\right)=\lambda\sum_{I=1}^{\Ncal}\beta^I\prod_{J\ne I}(\lambda-d_J)=\prod_{n=0}^{\Ncal-1}(\lambda-n\mu^2).
\end{align}
The last equality follows by polynomial interpolation: after removing the common factor $\lambda$, both sides have degree $\Ncal-1$ and agree at every $\lambda=d_I$. Hence the squared eigenmasses form an arithmetic progression,
\begin{align}
    \mu_n^2=n\mu^2,\qquad n=0,\ldots,\Ncal-1.
\end{align}
They are all distinct, and $\mu^2$ sets both the mass gap and the spacing of the squared masses.

More generally, replacing $n\mu^2$ in the interpolation formula by any prescribed distinct positive squared masses, and choosing the $d_I$ to interlace them strictly, gives positive coefficients $\beta^I$ with $\sum_I\beta^I=1$. The same choices of kinetic and cosmological parameters then realise that spectrum. Thus the arithmetic progression is an explicit example rather than a restriction imposed by the determinant interaction.

For the parameters above, an eigenvector for each eigenvalue can be chosen as
\begin{align}
    v^I_0=1,\qquad v^I_n=\frac{d_I}{d_I-\mu_n^2}=\frac{I-\tfrac12}{I-\tfrac12-n},\qquad n=1,\ldots,\Ncal-1.
\end{align}
The denominators never vanish, since the $d_I$ are half-integer multiples of $\mu^2$.

For profiles independent of the additional transverse coordinates $x^A$, the radial powers are
\begin{align}
    p_n^\pm=\tfrac12(D-1)\pm\nu_n,\qquad \nu_n=\sqrt{\tfrac14(D-1)^2+n\ell^2\mu^2}.
\end{align}
Imposing the AdS falloff $F_I=\mathcal O(z^{D-1})$ gives the general solution in this sector,
\begin{align}
    \label{distinct_mass_waves}
    F_I=a_{D-1}(x^+)\left(\frac{z}{\ell}\right)^{D-1}+\sum_{n=1}^{\Ncal-1}\frac{I-\tfrac12}{I-\tfrac12-n}\,a_n^+(x^+)\left(\frac{z}{\ell}\right)^{p_n^+}.
\end{align}
The first term is the common massless profile, while each term in the sum belongs to a different massive eigenmode.

For $D>3$, transverse dependence can be included by superposing nonzero Fourier modes,
\begin{align}
    f_n(x^+,z,\mathbf x)=\int\frac{\mathrm d^{D-3}\mathbf q}{(2\pi)^{D-3}}\,A_n(x^+,\mathbf q)e^{i\mathbf q\cdot\mathbf x}\left(\frac{z}{\ell}\right)^{(D-1)/2}I_{\nu_n}(qz),\qquad q=|\mathbf q|.
\end{align}
Whenever the integral and its derivatives converge, this solves the $n$th eigenprofile equation and has the required boundary falloff. Smooth coefficients of compact momentum support provide a sufficient choice. Reality is ensured by $A_n(x^+,-\mathbf q)=\overline{A}_n(x^+,\mathbf q)$. Combining these profiles with the zero-momentum solutions in \eqref{distinct_mass_waves} and reconstructing $F_I=\sum_n v^I_{n}f_n$ gives exact Siklos metrics without an expansion in the profile amplitudes.

\subsubsection{The pp-wave counterpart}
Keeping the same $c_I,\beta^I,m,m_I$ and replacing the bare cosmological constants by $\Lambda_I=-d_I$ gives a proportional Minkowski background. The mixing matrix, eigenvectors, and masses remain unchanged. The exact pp-wave profiles therefore take the form
\begin{align}
    F_I=H_0+\sum_{n=1}^{\Ncal-1}\frac{I-\tfrac12}{I-\tfrac12-n}\,f_n,\qquad \tilde\Delta H_0=0,\qquad (\tilde\Delta-n\mu^2)f_n=0.
\end{align}
Here $H_0$ is the common massless profile, while each $f_n$ belongs to a distinct massive eigenmode.

In $D=3$, these equations are ordinary differential equations in $y$, and their general solution is
\begin{align}
    F_I=a_0(x^+)+b_0(x^+)y+\sum_{n=1}^{\Ncal-1}\frac{I-\tfrac12}{I-\tfrac12-n}\left[a_n^+(x^+)e^{\sqrt n\mu y}+a_n^-(x^+)e^{-\sqrt n\mu y}\right].
\end{align}
The coefficient functions are arbitrary and real, and the common terms $a_0$ and $b_0y$ can be removed locally by a coordinate transformation.

For $D>3$, let $\mathbf x=(x^A)$ and Fourier transform in these additional transverse coordinates. The massive profiles admitting such a representation are
\begin{align}
    f_n(x^+,y,\mathbf x)=\int\frac{\mathrm d^{D-3}\mathbf q}{(2\pi)^{D-3}}\,e^{i\mathbf q\cdot\mathbf x}\left[A_n(x^+,\mathbf q)e^{\sqrt{q^2+n\mu^2}\,y}+B_n(x^+,\mathbf q)e^{-\sqrt{q^2+n\mu^2}\,y}\right].
\end{align}
The integrals are understood wherever they and their derivatives converge. The coefficients satisfy $A_n(x^+,-\mathbf q)=\overline{A}_n(x^+,\mathbf q)$ and $B_n(x^+,-\mathbf q)=\overline{B}_n(x^+,\mathbf q)$ for real profiles. Delta-function contributions at $\mathbf q=0$ reproduce the $x^A$-independent exponential solutions above. As in the general pp-wave construction, no global transverse boundedness is imposed.

The massless profile may be any real solution of $\tilde\Delta H_0=0$. For nonzero Fourier momentum its radial factors are $e^{\pm qy}$, while harmonic polynomials provide further simple examples. In $D=4$, for instance, $H_0=a(x^+)\bigl[y^2-(x^3)^2\bigr]$ gives a nontrivial massless pp-wave. Substituting these profiles into $F_I$ yields exact solutions at arbitrary amplitude. The change of background therefore alters the profile equations while preserving the prescribed mass spectrum and the eigenvectors that distribute each mode among the metrics.

\newpage
\section{Discussion}

We have constructed exact AdS waves and pp-waves in multi-gravity with a genuine multi-field interaction beyond pairwise couplings, for an arbitrary number of fields and in arbitrary spacetime dimension $D\geq3$. Our construction extends previous exact wave solutions in massive gravity \cite{Ayon-Beato:2009cgh,Mohseni:2011vv}, bimetric theory \cite{Mohseni:2012ug,Bergshoeff:2014eca,Ayon-Beato:2018hxz}, and multi-gravity with pairwise interactions \cite{Garcia-Compean:2026cnq,Ayon-Beato:2026xaz}. In particular, the recent work on multi-gravity considers a pairwise chain of nearest-neighbour interactions. In this work, the full determinant interaction of \cite{Hassan:2018mcw} couples all the fields and thus the wave profiles. After diagonalising the field equations, the wave profiles obey differential equations with the Fierz--Pauli masses as mass parameters. This connection, already established in bimetric theory \cite{Ayon-Beato:2018hxz}, is extended here to the complete mass spectrum of the genuine multi-field interaction \cite{Flinckman:2024zpb}. We also demonstrated that the determinant interaction admits wave configurations unavailable to standard pairwise tree theories with the same number of fields, thus providing new multi-gravity solutions.

A useful feature of this interaction is the algebraic control over its spectrum. The existence results for proportional solutions on the branch $\beta^Ic_I>0$ \cite{Flinckman:2024zpb}, together with the bounds \eqref{sandwitch}, identify a branch with non-negative squared eigenmasses without requiring explicit roots of the characteristic polynomial. The product of the remaining eigenvalues further shows that an additional massless mode cannot arise at finite regular parameters with positive kinetic weights and nonzero interaction coefficients.\footnote{Additional zero-mass profiles have been considered in bimetric wave solutions \cite{Ayon-Beato:2018hxz}, while critical massless degeneracies have been studied in Zwei-Dreibein, Viel-dreibein and higher-derivative gravities \cite{Bergshoeff:2014eca,Ayon-Beato:2026xaz,Bergshoeff:2012ev,Nutma:2012ss}. These involve different interaction potentials or kinetic regimes. At critical points, repeated massless roots can correspond to logarithmic generalised modes rather than independent massless Fierz--Pauli fields. Their nonlinear consistency requires separate analysis; for example, the proposed unitary truncation of parity-even tricritical gravity suffers from a linearisation instability \cite{Apolo:2012vv}.} On AdS solutions, the imposed boundary falloff, which reduces to the Brown--Henneaux condition in ${D=3}$, admits the appropriate radial branch of every eigenmode with non-negative squared mass. On other branches, it excludes profiles with negative squared masses, although this restriction on the allowed excitations does not by itself exclude the corresponding solutions.

Although the massless eigenvalue is unique, the massive eigenvalues can coincide. With positive kinetic weights, the profile mixing matrix $\Mcal$ is similar to a real symmetric matrix and therefore admits a complete eigenbasis even at such degeneracies. The same assumption ensures healthy positive kinetic terms for the linearised spin-2 modes. Coincident massive eigenvalues consequently do not generate the additional logarithmic solutions encountered at critical points of Zwei-Dreibein and Viel-dreibein gravity \cite{Bergshoeff:2014eca,Ayon-Beato:2026xaz}. This reasoning should also apply to pairwise theories whenever their mixing matrices and kinetic terms satisfy the same condition. Logarithms arising from repeated radial roots at the Breitenlohner--Freedman threshold have a different origin and are excluded here by the imposed boundary falloff.

Our ansatz describes common null wavefronts $x^+=\mathrm{constant}$ with profiles independent of $x^-$. This restricted dependence does not determine the group velocities of massive wave packets. Studying dispersion and observable gravitational propagation requires more general perturbations and a specified matter coupling. The pp-wave sector illustrates another consequence of the restricted ansatz. For $\mu_n^2>0$, the transverse equation ${(\tilde\Delta-\mu_n^2)f_n=0}$, whose counterparts also arise in massive gravity and bimetric theory \cite{Mohseni:2011vv,Ayon-Beato:2018hxz}, admits no nonzero smooth, bounded profile on the entire transverse space. This is a restriction on globally bounded, source-free solutions within the ansatz and does not signal an instability of the massive modes. A natural extension is to include matter coupled to a single metric, with its stress tensor aligned with the common null direction, and study how it sources the different eigenmodes. Scalar and electromagnetic sources have already been investigated for bimetric AdS waves \cite{Ayon-Beato:2018hxz}.

The shared Kerr--Schild structure also connects these solutions to other exact configurations. A common null deformation simplifies the interaction terms in constructions of rotating bimetric black holes \cite{Ayon-Beato:2015qtt} and multi-metric black holes \cite{Wood:2024acv,Wood:2024eol}. It would be useful to determine how far this approach extends to other geometries in the determinant theory. Related Kerr--Schild constructions have also been used to study the classical double copy in bimetric theory and pairwise multi-gravity \cite{Garcia-Compean:2024uie,Garcia-Compean:2026cnq}. The present solutions provide a setting in which to investigate whether genuine multi-field interactions admit a corresponding description in terms of coupled gauge and scalar fields.

The arbitrary dependence on $x^+$ permits wave pulses supported within a finite interval of this coordinate. Such solutions could be used to investigate gravitational memory by comparing geodesic separations and relative velocities before and after the pulse, following the analyses of exact plane waves in General Relativity \cite{Zhang:2017rno,Chakraborty:2019yxn}. In multi-gravity, this requires specifying the metric followed by the test particles. The resulting tidal response could then reveal how the different massive eigenprofiles contribute to displacement or velocity memory.

Several questions about the physical solution space remain open. In particular, the AdS boundary falloff does not establish regularity at the Poincar\'e horizon, and positive Fierz--Pauli masses do not establish stability of the finite-amplitude wave backgrounds. Their global regularity, perturbative stability, and conserved charges require separate analyses.

An extension to de Sitter space would complete the maximally symmetric cases, but requires a more general Kundt geometry without the null Killing symmetry of the Siklos ansatz \cite{Ortaggio:2024say}. Related Kundt waves on AdS have already been constructed in pairwise multi-gravity \cite{Garcia-Compean:2026cnq}. In dS, the massive spectrum would also have to satisfy the Higuchi bound \cite{Higuchi:1986py}, so positivity of the squared masses alone would be insufficient.

The results presented here establish an exact gravitational-wave sector of genuine multi-gravity in which the Fierz--Pauli spectrum governs finite-amplitude profiles. The massless and massive contributions identified at quadratic order thus admit an explicit realisation in solutions of the full nonlinear field equations, extending the construction of exact gravitational waves beyond pairwise interactions.

\newpage

\bibliographystyle{JHEP}
\bibliography{biblio.bib}

\end{document}